\documentclass[a4paper]{elsarticle}
\pdfoutput=1 

\usepackage{lineno}
\usepackage[utf8]{inputenc} 
\usepackage{amsmath,amssymb,amsfonts}

\usepackage{soul}
\usepackage[dvipsnames]{xcolor}
\usepackage{graphicx}
\usepackage[version=3]{mhchem}	

\journal{Nuclear Physics A}

\begin{document}

\begin{frontmatter}

\title{An automated system to define the optimal operating settings of cryogenic calorimeters}

\author[c]{K.~Alfonso}
\author[d]{C.~Bucci}

\author[d,e]{L.~Canonica \corref{cor2}}
\cortext[cor2]{Currently at Max-Planck-Institut f\"ur Physik, D-80805 M\"unchen, Germany}

\author[a,b]{P.~Carniti}
\author[f,g]{S.~Di Domizio}
\author[a,b]{A.~Giachero}
\author[a,b]{C.~Gotti}
\author[h,i]{L.~Marini}

\author[a,b]{I.~Nutini\corref{cor1}}
\cortext[cor1]{Corresponding author}
\ead{irene.nutini@mib.infn.it}

\author[a,b]{G.~Pessina}

\address[a]{Dipartimento di Fisica, Universit\`{a} di Milano-Bicocca, Milano I-20126, Italy}
\address[b]{INFN -- Sezione di Milano Bicocca, Milano I-20126, Italy}
\address[c]{Department of Physics and Astronomy, University of California, Los Angeles, CA 90095, USA}
\address[d]{INFN -- Laboratori Nazionali del Gran Sasso, Assergi (L'Aquila) I-67100, Italy}
\address[e]{Massachusetts Institute of Technology, Cambridge, MA 02139, USA}
\address[f]{Dipartimento di Fisica, Universit\`{a} di Genova, Genova I-16146, Italy}
\address[g]{INFN -- Sezione di Genova, Genova I-16146, Italy}
\address[h]{Department of Physics, University of California, Berkeley, CA 94720, USA}
\address[i]{Nuclear Science Division, Lawrence Berkeley National Laboratory, Berkeley, CA 94720, USA}

\begin{abstract}
Cryogenic macro-calorimeters instrumented with NTD thermistors have been developed for several decades.
The choice of the optimal bias current is crucial for a proper operation of these detectors, both in terms of energy resolution and stability.
In this paper we present a set of automatic measurements and analysis procedures for the characterization and optimization of the working configuration of
the NTD thermistors.
The presented procedures were developed for CUORE, an array of 988 cryogenic macro-calorimeters instrumented with NTD thermistors that has been taking data
since 2017.
These procedures made it possible to characterize a large number of detectors in a reliable way. They are suitable enough to be used also in other
large arrays of cryogenic detectors, such as CUPID.
\end{abstract}



\end{frontmatter}


\section{Introduction}\label{sec:intro}

Cryogenic calorimeters 
represent one of the leading techniques for rare event searches.
This class of low temperature detectors features an excellent energy resolution, low energy threshold and a wide choice of absorber materials~\cite{bolometers:enss}.
In particular, the energy resolution and threshold depend on the sensor, the mass and the material of the absorber, and on the application-specific configuration of the detector.
For large cryogenic calorimeters, the energy resolution is on the order of few keV, close to that of the germanium semiconductor detectors, and the threshold can be as low as a few keV \cite{Alduino:2017xpk}.

At present, the largest implementation of the low temperature calorimetric technique is CUORE (Cryogenic Underground Observatory for Rare Events), an experiment for the search for $^{130}$Te neutrinoless double beta decay (0$\nu\beta\beta$) \cite{FIORINI:1998309,Artusa:2014lgv}. The use of cryogenic calorimeters for investigating double-beta decay has a long history of development \cite{BIASSONI2020103803}. In these last 30 years, experiments utilizing TeO$_2$ crystals grew in mass and complexity \cite{doi:10.1063/1.5031485}. Successful demonstrators leading the path forward to CUORE were MiDBD \cite{ARNABOLDI:2003167}, Cuoricino \cite{ANDREOTTI:2011822}, CUORE-0 \cite{CANONICA:201573,Alduino:2016vjd}. 
The CUORE experiment is located at the Gran Sasso National Laboratories of INFN in central Italy, and has been taking data since 2017. 
The detector is an array of 988 TeO$_2$ crystals operated as cryogenic calorimeters, with a total mass of 741 kg. It is hosted in a custom cryogenic infrastructure that keeps the array at a temperature of $\sim$10\,mK \cite{Alduino:2019xia,DAddabbo:2018jjw}.
CUORE was constructed to have an energy resolution of 5 keV FWHM in the region around the Q-value of the 0$\nu\beta\beta$ decay (2527 keV for $^{130}$Te).

\begin{figure}[t!]
\centering
\includegraphics[width=0.9\textwidth]{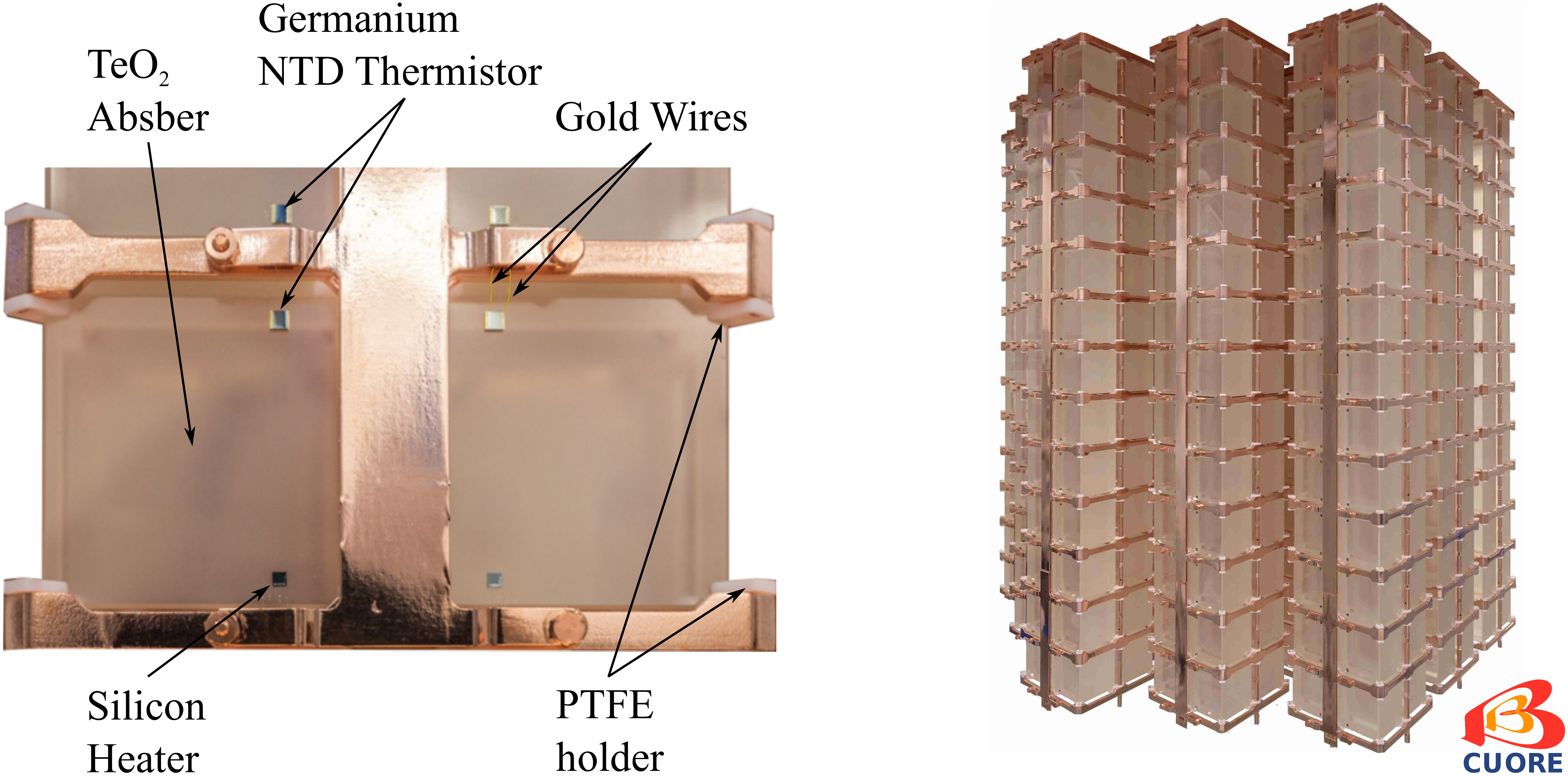}
\caption{[Left] CUORE instrumented detectors. A single unit consists of: TeO$_2$ crystal, NTD thermistor, silicon heater, readout wires connected to copper bands and PTFE holders. [Right] The CUORE detector: the 19 tower modules hosting the 988 TeO$_2$ crystals.~\cite{Nutini:2020vtd}} 
\label{fig:cuore}
\end{figure}

A single CUORE detector is composed of a 5$\times$5$\times$5 $\mbox{cm}^3$ TeO$_2$ absorber crystal with a mass of 750\,g, and a 3$\times$3$\times$1 $\mbox{mm}^3$ neutron transmutation doped (NTD) germanium thermistor that is thermally coupled to the crystal by means of glue spots and used for the temperature readout \cite{Alduino:2016vjd}. The CUORE NTD thermistors are germanium chips doped with thermal neutrons and they act as phonon sensors \cite{Larrabee:1984}. The small size of the NTD thermistor compared to the crystals limits its contribution to the detector heat capacitance, while ensuring good electrical properties in terms of resistance value. 
Each crystal is also instrumented with a silicon heater to periodically inject a fixed amount of energy to the detector for gain stabilization \cite{ANDREOTTI2012161,Alessandrello:1998}.
The crystals are held by a copper structure that is in thermal equilibrium with the coldest point of the dilution refrigerator, and behaves as a heat sink at $\sim$10\,mK. The mechanical coupling between the crystals and the copper frames is made by PTFE supports, while the thermal link towards the heat sink constitutes of gold wires used for the readout of the NTD thermistors. 
A picture of the CUORE instrumented detectors is shown in fig.\,\ref{fig:cuore}[left]. The CUORE TeO$_2$ crystals are arranged into 19 identical structures called towers. Each tower hosts 52 detectors arranged in 13 floors, each containing 4 crystals. See fig.\,\ref{fig:cuore}[right].

 The NTD thermistor converts the thermal pulse, produced by an energy release in the absorber crystal, into a resistance variation that is proportional
to the released energy.
The NTD thermistor is inserted in a bias circuit that converts the resistance variation into a voltage signal.
The determination of the optimal NTD thermistor bias current is crucial to obtain the best performance from the detector. The behavior of the detector with respect to the applied bias is described by the characteristic current-voltage (I-V) curve.
The I-V curve measurement and the identification of the optimal bias current, or optimal working point (WP), is a standard procedure for the operation of cryogenic calorimeters.
CUORE's predecessor experiments 
utilized only a few tens of detectors.  
In those cases the characterization measurements were performed by manually changing the configuration of the bias circuit and recording the corresponding output voltage.
Moreover there was no automated analysis procedure, and the working point was chosen from a visual inspection of the measured curves. 
In CUORE, the large number of calorimeters and the intrinsic spread of their characteristics, due to slight non-uniformities of both crystals and thermistors, necessitated the development of new automatic procedures for the characterization and optimization of various detector parameters~\cite{Nutini:2020vtd,Dompe:2020JLTP,DAddabbo:2017efe}. For a large number of calorimeters, an automated procedure allows faster and more precise measurements and a better control over the detectors performance.
In addition, the optimal working points of the NTD thermistors were identified by an automated analysis procedure which takes into consideration not only the I-V curve and the voltage amplitude, but also the pulse shape variation with the bias current. 

In the following we will describe these procedures in more detail.
We will first recall in sec.\,\ref{sec:lc_ntd} the working principle of the NTD thermistor and the parameters that play major roles in the characterization of these devices.
In sec.\,\ref{sec:lc_meas}, we will give an overview of the hardware and slow control software, with which the data from the CUORE detectors are controlled and acquired.
We will provide a detailed description of the aforementioned procedures for characterizing the NTD thermistors in the last part of sec.\,\ref{sec:lc_meas}. sec.\,\ref{sec:lc_analysis} will focus on the developed analysis algorithms for identifying the optimal working point. In that same section, we will show the viability of this procedure in optimizing the NTD thermistors operating conditions at different temperatures.

The procedures described in this paper were applied to the CUORE experiment, setting the foundation for its scientific results \cite{Alduino:2017ehq, Adams:2019jhp}. Moreover, the CUPID-0 experiment \cite{Azzolini:2018tum}, a demonstrator of the scintillating calorimetric technique for the search of $^{82}$Se neutrinoless double beta decay, also profited from this automated system for measuring the NTD characteristic curves. This allowed the setting of optimal working conditions for the NTD thermistors coupled with the ZnSe crystals, and the achievement of several physics results \cite{Azzolini:2020skx,Azzolini:2019yib,Azzolini:2019tta,Azzolini:2018oph,Azzolini:2018dyb}.
Given the positive applications of the procedure, the same automated system can be applied to any array of cryogenic calorimeters read by NTD thermistors, such as the CUORE future upgrade, CUPID \cite{CUPID-pre-CDR:2019}.

\section{NTD thermistors characterization}\label{sec:lc_ntd}

\noindent The NTD thermistors convert thermal pulses into electrical signals through a resistance variation. They are sensitive only to thermalized phonons in the absorber, thus acting as ideal thermometers.
The NTD thermistors are semiconductor (usually germanium) slabs doped by means of thermal neutrons \cite{Larrabee:1984,PhysRevB.41.3761}.
When the doping concentration reaches a critical value, the semiconductor enters the metal-insulator transition (MIT) region, where its resistivity exhibits a strong dependence on the temperature \cite{MottDavies:1980}.
At sub-Kelvin temperature, the thermistor resistance dependence on the temperature $R_{NTD}(T)$ is described by  \cite{Miller:1960vrh}:
\begin{equation}\label{eq:R_NTD}
R_{NTD}(T) = R_0 \text{ exp}{\left(\cfrac{T_0}{T}\right)^{\gamma}}
\end{equation}
where $R_0$ and $T_0$ depend on the doping concentration and on the geometry, $T$ is the temperature, and $\gamma$ depends on the conduction mechanism.
The parameters $R_0$, $T_0$ and $\gamma$ can be estimated by measuring the thermistor resistance at various temperatures.
For CUORE NTD thermistors, typical values are: $R_0$\,=\,1.0\,--\,1.5\,$\Omega$, $T_0$\,=\,4.0\,--\,5.0\,K , and $\gamma$\,$\simeq$\,0.5\,\cite{Alduino:2016vjd}.
At T $\sim$ 10 mK, typical resistance values for CUORE NTD thermistors are of the order of a few hundred ~M$\Omega$.  An energy deposition of 1\,MeV in the absorber produces a temporary increase of the temperature of the order of a few hundred\,$\mu$K, inducing a resistance variation of a few M$\Omega$. 

\subsection{Operating NTD thermistors and load curves}
\begin{figure}[t!]
\centering
\includegraphics[width=\textwidth]{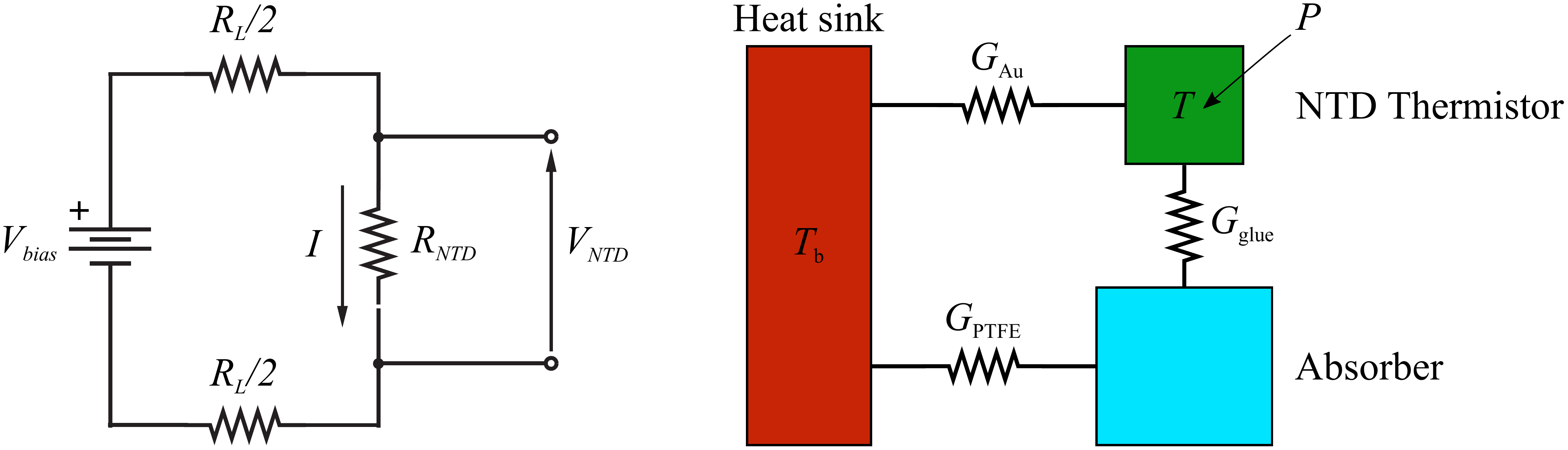}
\caption{[Left] Scheme of the biasing circuit for the thermistor readout. [Right] Simplified scheme of a thermal detector. The absorber and the NTD thermistor are modeled as a system weakly coupled to the heat sink, via both the PTFE and the gold wires (Au) conductances.}
\label{fig:boloBias}
\end{figure}

\begin{figure}[t!]
\centering
\includegraphics[width=0.7\textwidth]{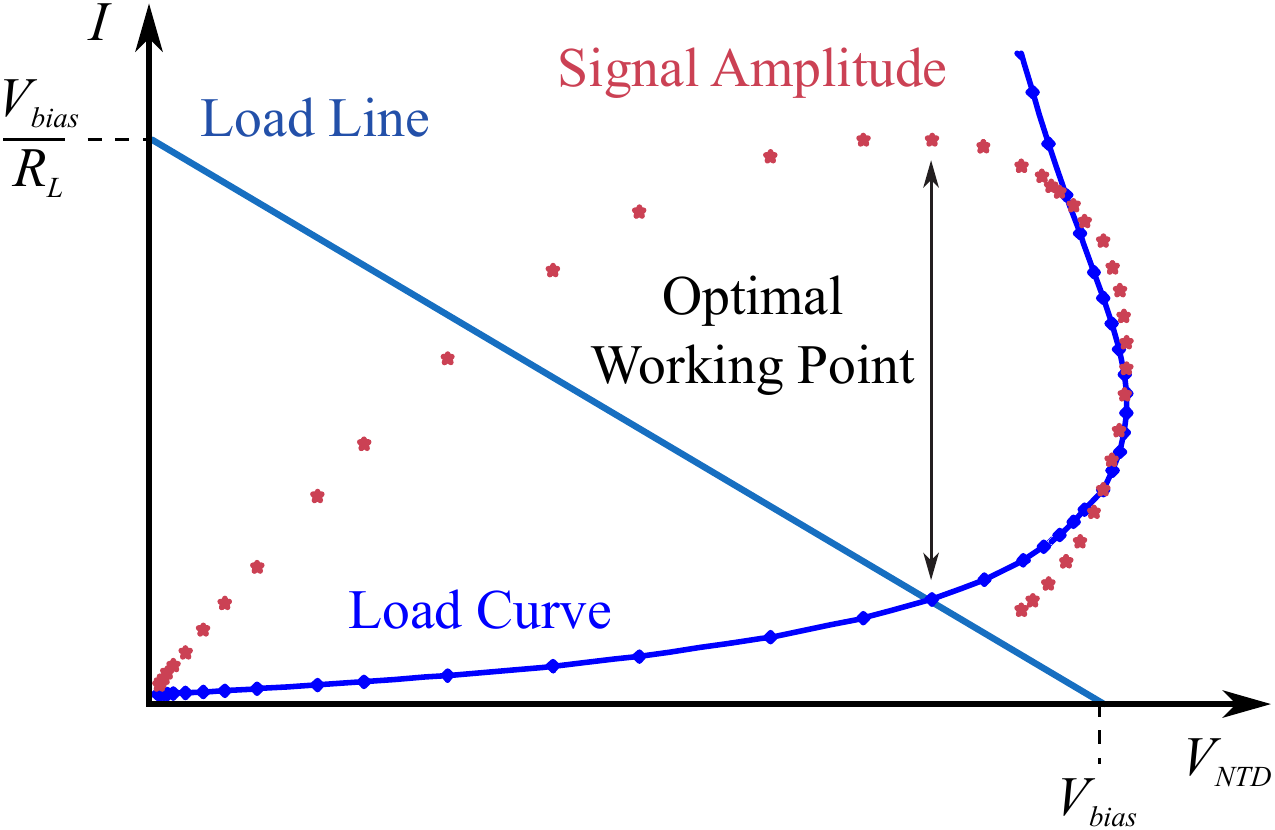}
\caption{Load curve (I-V) of an NTD thermistor [blue dots/line], {an analytical representation} of the static behavior. Superimposed to the former curve, there is the plot of the variation of pulse amplitude A for increasing bias current for the same injected energy [red dots]; this is an example of the dynamic behavior of an NTD thermistor.} 
\label{fig:boloBias1}
\end{figure}

To measure the NTD thermistor resistance and its variation, the sensor is biased with the circuit schematically represented in fig.\,\ref{fig:boloBias}[left]. 
A bias voltage $V_{bias}$ is applied across a pair of load resistors $R_L/2$, in series with the thermistor.
The total resistance $R_L$ is chosen much higher than the thermistor resistance $R_{NTD}$, so that the current I flowing through the thermistor can be considered approximately constant in the static condition.
The voltage across the thermistor and the power dissipation are:
\begin{equation}\label{eq:IV}
V_{NTD} = I \cdot R_{NTD}(T) \quad , \quad P = I^2 \cdot R_{NTD}(T)
\end{equation}
The thermistor temperature $T$ is affected by the power dissipation P: 
\begin{equation}
T = T_b + \frac{P}{G}
\end{equation}
where $T_b$ is the temperature of the heat sink and
$G$ is the conductance between the NTD thermistor and the heat sink, as shown in the simplified model in fig.\,\ref{fig:boloBias}[right]. 
For high power dissipation, the increase in the thermistor temperature P/G is comparable to $T_b$ and is not negligible. 
The power dissipation heats up the NTD thermistor resulting in a smaller resistance, according to eq. \ref{eq:R_NTD}.
This phenomenon is called electro-thermal feedback and it will be discussed in more detail in sec \ref{sec22}.
The I-V curve for semiconductor thermistors is usually referred to as the 
load curve.
An example of a load curve is shown in fig.\,\ref{fig:boloBias1}.
Given a fixed value of $T_b$, the slope of the load curve is nearly constant for low I.
When the bias current is sufficiently high, the electro-thermal feedback causes the I-V relation to deviate from linearity leading to a non-ohmic behavior. 
Each point on the I-V curve is a potential WP.
As shown in fig.\,\ref{fig:boloBias1}, a WP is the point of intersection between the load line: 
\begin{equation}
V_{NTD} = V_{bias} - I \cdot R_L 
\end{equation}
and the load curve.
We change the NTD thermistor bias current, in order to find the detector's best operating conditions along the I-V curve, i.e. the optimal WP. The NTD thermistor dynamic response to thermal pulses also plays a key role in the choice of the optimal WP.

\subsection{Response of a NTD thermistor to thermal pulses}\label{sec22}
For cryogenic calorimeters, an energy release in the absorber causes a temperature variation that is converted to a voltage pulse by the NTD thermistor and its biasing circuit.
The amplitude of the voltage signal depends on the specific detector configuration and biasing circuit. In CUORE, it has an amplitude of $\sim$ 100 -- 400 $\mu$V/MeV and requires amplification.

For signal pulses, the relationship between the maximum voltage variation (or pulse amplitude), A, and the energy deposition resulting in a temperature variation, $\Delta T$, is:
\begin{equation}\label{eq:ampli_t}
A \propto \frac{\Delta T}{T}\,V_{NTD}
\end{equation}
This dependence holds if the thermistor logarithmic sensitivity \cite{Stahl:2005} is used to correlate resistance and temperature variations.
Eq. \ref{eq:ampli_t} shows that, for a reference energy release, its pulse amplitude is proportional to $V_{NTD}$ and to the relative temperature variation $\Delta T/T$. 
For low values of $V_{bias}$, when the electro-thermal feedback is negligible, the temperature of the NTD thermistor is almost independent of $V_{bias}$.
In this regime the pulse amplitude A increases for increasing applied bias, since $V_{NTD}$ $\propto$ $V_{bias}$ in this regime. At higher bias values, when the electro-thermal feedback comes into play, the NTD thermistor starts to heat up, causing a reduction in $\Delta T/T$. Therefore at some point, as the electro-thermal feedback becomes more and more important, the A vs $V_{bias}$ curve will reach a maximum before decreasing, as shown in fig.\,\ref{fig:boloBias1}.
The electro-thermal feedback also has an effect on the pulse shape; the non-ohmic behavior of the device for high bias current leads to instabilities in its dynamic response, resulting in distorted pulses or in an oscillation of the detector.
In order to avoid regenerative electro-thermal feedback, semiconductor thermistors should not be loaded with an impedance smaller than the impedance of the thermistor itself.
In CUORE, the NTD thermistors are read out by an amplifier at room temperature and this leads to a large parasitic capacitance of the link in parallel to the sensor.
This capacitance gives the dominant contribution to the load impedance in the bandwidth of interest. 
At low bias, the NTD thermistor impedance is purely resistive and equals the ratio of its bias voltage and current: in this case an oscillation can start if the signal has a frequency content for which the module of the parasitic capacitance becomes smaller than that of the thermistor impedance.
The feedback, however, increases the power and the NTD thermistor lowers its resistance, damping the oscillation.   
At high bias, the NTD thermistor deviates from the resistive regime and its dynamic impedance shows an inductive component: in this case, when the inductive component gets sufficiently high, the oscillation is maintained instead of being damped \cite{Arnaboldi:2005}.
This region of operation must be avoided in order to have a stable detector operation.

The optimal WP for a NTD thermistor is chosen as a compromise between having a large signal-to-noise ratio (SNR) and ensuring a stable behavior of the device. The SNR is evaluated as the ratio of the pulse amplitude and the noise level of the detector for each bias configuration. In order to measure the pulse amplitude and shape variation with bias voltage, a reference energy injection is needed.
For this purpose, either a line from a reliable radioactive source, or a given power injection from the heater installed on the crystal, could be used.
{For large low temperature calorimeters, the two approaches are almost equivalent, since we can make the shape of heater events resembling the one of particle events.}
Generally, the latter solution is chosen because of the higher precision of the energy injection and the possibility of regulating the rate of heater pulses for each NTD thermistor bias configuration. 
{Moreover, while the resolution for physics events increases monotonically with energy, the SNR of heater events is compatible with the lower achievable resolution, dominated by residual noise on the detectors, and it is therefore a good proxy for the studies of the detector SNR variation with bias voltage.}
The noise level  of the detectors could be theoretically calculated, considering the NTD thermistor readout and amplification circuit. The intrinsic noise is expected to be dominated by the parallel Johnson noise of the load resistances across the impedance of the NTD thermistor, which is proportional to its resistance and to the ratio of the temperatures of the load resistors and the NTD thermistor. The amplifier stage is usually designed such that its series noise is negligible. However, there can be further contributions to the noise related to vibrations, pick-up noise, etc. which can be difficult to model and quantify for different detector setups. Therefore, we evaluate the noise level variation for each bias configuration directly from the analysis of the fluctuations of the NTD output voltage.

\section{Performing characterization measurements}\label{sec:lc_meas}
In real experimental conditions, we do not have `a priori' knowledge of all the NTD thermistor and detector setup parameters. Moreover, in case of experiments like CUORE with a large number of cryogenic calorimeters, every detector is not exactly the same and it is necessary to optimize each crystal's NTD thermistor bias voltage.
Therefore the NTD thermistor optimal WP is determined experimentally and dedicated measurements on the NTD thermistors are necessary. 
Three types of measurements, which we will refer to as {\it{characterization measurements}}, can be performed:
\begin{itemize}
\item {\it{Load curve measurement:}} accurately reconstruct the I-V curve, in order to characterize the NTD thermistor static behavior and identify the range of bias voltages for which the pulse amplitude is close to its maximum;
\item {\it{Working point measurement:}} accurately build the SNR curve and study the pulse shape variation as a function of the applied bias voltage in order to find the optimal WP;
\item {\it{Resistance measurement:}} precisely measure the NTD thermistor static resistance at a given bias voltage to monitor its stability over time.
\end{itemize}
Estimators for the NTD thermistor resistance, voltage variation for a given energy deposition, and noise are therefore calculated for each different bias current value from the mentioned measurements.
The procedures and algorithms used to reconstruct the NTD thermistor's voltage, current and resistance, noise, pulse amplitude, signal-to-noise ratio will be described in detail in sec.\,\ref{sec:lc_analysis}.

\begin{figure}[t!]
\centering
\includegraphics[width=0.95\textwidth]{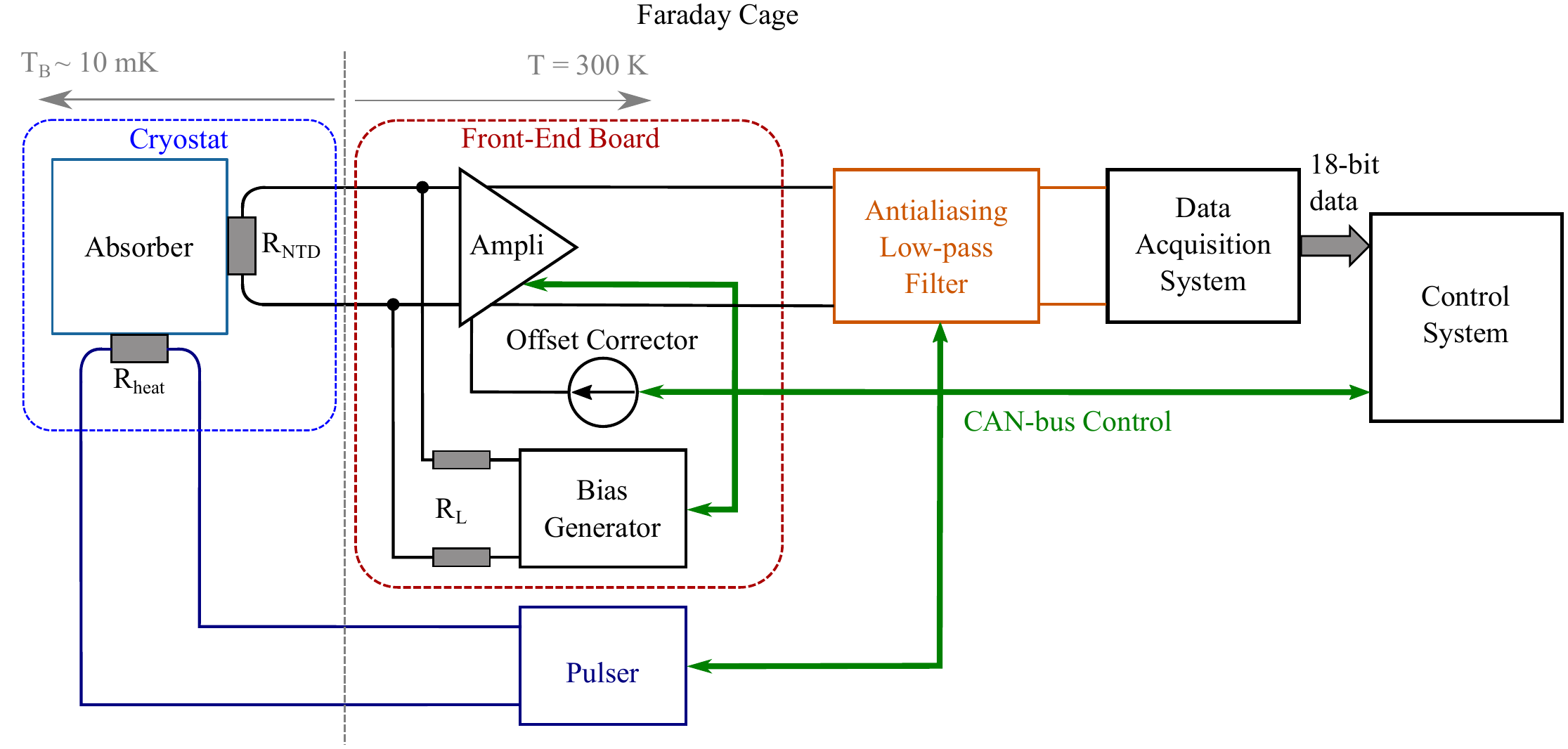}
\caption{Schematic of the CUORE detectors electronics and DAQ chain.} \label{fig:Ele_DAQ_simple}
\end{figure}

A dedicated procedure has been developed in order to perform the characterization measurements of the response parameters for each of the 988 CUORE detectors.
In the following, we will first present {in sec.\,\ref{sec:ele}} the CUORE electronics systems for the readout of the NTD thermistor and for power injection through the silicon heater (${R_{heat}}$), as represented in fig.\,\ref{fig:Ele_DAQ_simple}.
We will focus on the design specifications of the CUORE front-end boards, utilized for the NTD thermistor biasing circuit (introduced in sec.\,\ref{sec:lc_ntd}) and amplification stage, and the capabilities of the pulser board. {In sec.\,\ref{sec:daq}}, the function of the CUORE data acquisition system (DAQ) to identify and associate raw data from the detector with designated electronics settings will be discussed in relation to the {\it{characterization measurements}} on the NTD thermistors.
Lastly, we will describe {in sec.\,\ref{sec:lc_processing}} the load curve measurement, including the front-end and DAQ configurations for the procedure.
We emphasize the design considerations of each infrastructure to facilitate the operation of a large array of cryogenic detectors.

\subsection{Electronics}\label{sec:ele}
The front-end electronics system, specifically developed for the CUORE experiment, is responsible for the differential signal amplification, detector biasing, and detector thermal response calibration and stabilization.
The first two functions are performed by the JFET-input preamplifier and the front-end board~\cite{FrontendCUORE}, while the calibration is handled by the pulser board~\cite{PulserCUORE}.
Auxiliary electronics systems provide anti-aliasing filtering~\cite{BesselCUORE}, power supply regulation, and high-stability voltage references~\cite{ACDCCUORE, LinearSupplyCUORE}.
The technical details of these electronics systems are thoroughly described in the references provided above; in the following we will highlight the characteristics that influence the {\it characterization measurement}.

The detector bias current is generated by applying an adjustable differential voltage to two alternative pairs of high-value custom-made load resistors placed at room temperature, which have a designated value of ${60\:\mbox{G}\Omega}$ (${2\times30\:\mbox{G}\Omega}$) or ${10\:\mbox{G}\Omega}$ (${2\times 5\:\mbox{G}\Omega}$).
For this reason, in the following sections we use the term bias voltage instead of bias current.
The resistor value selection is done through relays, whose layout was optimized to avoid thermal drifts of the bias current \cite{FrontendCUORE}.
Having two possible values for the load resistors enables proper readout of a broad range of detector impedance, from a few hundred k$\Omega$ up to a few G$\Omega${, so that the same electronics can be used for very different experimental conditions (base temperature, bias voltage, and NTD type). The spread in the CUORE detectors' impedance at the optimal working point, as will be shown in sec.\,\ref{sec:wp_check}, is much lower.}

The bias voltage has a maximum differential range of $\pm$50~V, which corresponds to 0.8~nA (5~nA) bias current with ${30\:\mbox{G}\Omega}$ (${5\:\mbox{G}\Omega}$) load resistors, a resolution of 16 bits, and a thermal drift better than 80 ${ppm/^\circ C}$.
Each of the two differential voltages (positive and negative) can be set independently from the other, providing the capability of applying a bias voltage with non-zero common-mode voltage.
This is useful when evaluating and optimizing the DC common-mode rejection ratio (CMRR) of the preamplifiers, which is important for suppressing any common-mode noise source; internal trimmers are used for optimizing the CMRR.
Once the bias voltage is set by the on-board circuitry, a relay is used to invert the polarity of the bias applied to the detector without further adjustments.
This operation provides the capability of taking the difference between the two configurations (direct and inverted bias), thus canceling purely electrical offsets due to the preamplifier voltage offset.
In the case when the load curves are acquired with a single polarity, the effect of the offset can be minimized by adjusting it to zero before performing the measurements.
The JFETs' gate current (below 20~fA differential and below 100~fA single-ended) was designed to be completely negligible with detector impedance up to a few ${\mbox{G}\Omega}$~\cite{FrontEndInputCUORE}.
If the gate current were not negligible, its residual power injection could still be taken into account~\cite{ALESSANDRELLO199727}.
The signal amplification is fully differential and the gain is remotely programmable (from 20~V/V up to 10k~V/V) in order to accommodate the broad dynamic range of this type of detector when operated at different possible base temperatures.
The offset can be digitally adjusted up to ${\pm 80\ mV}$ (input-referred) and its stability is calibrated to be better than ${1\:\mu\mbox{V}/^\circ\mbox{C}}$, enabling reliable baseline monitoring even for very long operating times (years).

Detector thermal response stabilization is performed using a pulser board that is capable of applying precise (sub ${\mbox{ppm}/^\circ\mbox{C}}$), low-noise voltage pulses to a Si heater resistor glued to the detector.
The response of the detector to such a heat pulse enables the monitoring of the detector behavior at different base temperatures and bias currents.

All the adjustments done by the front-end electronics (bias voltage, relay switching, offset adjustment, pulse generation, etc.) are managed by the on-board firmware.
In this way, each operation can be easily parallelized since the DAQ provides only a few slow control commands to the front-end electronics through the CAN-bus interface, while all the computation algorithms are managed by the on-board microcontrollers.

The specifications and characteristics of the front-end systems are summarized in tab.\,\ref{tab:ele_specs}.

\begin{table}
    \centering
    \begin{tabular}{r|l}
        Signal amplification        & Fully differential \\
        Input transistors           & InterFET NJ132 (custom pairs) \\
        Gain settings               & 20\:V/V -- 10k\:V/V \\
        Gain stability              & $<5\:\mbox{ppm}/^\circ\mbox{C}$ \\
        Offset adjustment           & ${\pm 80\:\mbox{mV}}$ \\
        Offset stability            & ${<1\:\mu\mbox{V}/^\circ\mbox{C}}$ \\
        Input bias current          & ${<100\:\mbox{fA}}$ \\
        Input differential current  & ${< 20\:\mbox{fA}}$ \\
        Series noise                & 3.3\:nV/${\sqrt{\mbox{Hz}}}$ (white) \\
                                    & 8\:nV/${\sqrt{\mbox{Hz}}}$ (1\:Hz) \\
        Parallel noise              & $<0.15\:\mbox{fA}/\sqrt{\mbox{Hz}}$ \\
        CMRR (after optimization)   & ${> 120\:\mbox{dB}}$ \\
        Bias voltage $\left(V_{bias}\right)$ range      & $\pm$50\:V \\
        Load resistors $\left(R_L\right)$               & ${2\times30\:\mbox{G}\Omega}$ or ${2\times5\:\mbox{G}\Omega}$ \\
        Bias current range          & $\pm$0.8\:nA @ 60\:G${\Omega}$, $\pm$5\:nA @ 10\:G${\Omega}$ \\
        Bias voltage resolution & 16 bits \\
    \end{tabular}
    \caption{Summary of the front-end electronics specifications.}
    \label{tab:ele_specs}
\end{table}

\subsection{DAQ}
\label{sec:daq}
The CUORE data acquisition system, described in detail in~\cite{DiDomizio:2018ldc}, consists of a core system for data digitization and storage, and of a control interface for the analog electronic readout chain.
The detector waveforms are acquired with a 1\:kHz sampling frequency by a 18-bit ADC.
The relatively low signal bandwidth (on the order of 10 Hz) makes it possible to digitize and save the continuous waveforms of the detectors for offline processing.
For the same reason, triggering can be performed in software, allowing for more flexibility and control.
The data analysis is performed on triggered events, namely finite-length waveform windows selected in correspondence of a trigger.
The default event window length used in CUORE is 10\:s.
Events are associated with supplementary information including {DetectorId} (an identifier which maps the detectors in the CUORE array), time, and the type of trigger that caused the generation of the event. 
Three types of events exist: {signal events}, generated when the signal trigger detects a pulse in the waveform; {noise events}, generated periodically or at a random time, regardless of the presence of a pulse in the waveform; {pulser events}, whose generation is forced in correspondence to the injection of a pulse from a heater attached to the absorber. 

The control software for interfacing with the analog electronics is implemented with a multi-threading approach, allowing concurrent communication with multiple targets using a single CAN-bus link.
Some digital signals are acquired synchronously with the detector waveforms and are used to synchronize external events with the data.
In particular, digital signals are generated in correspondence to changes in the electronics configuration and when pulser-generated signals are injected in the detectors.
The synchronization mechanism enables pulser-generated events to be identified and flagged in the offline analysis.
Similarly, each acquired event can be associated with an EleId, i.e. an identifier for a well-defined electronics configuration.
This capability of characterizing the measurement procedure is fundamental for the offline analysis because it enables events to be classified based on their EleId.
Quantities related to the steady-state performance of the detector are estimated from the noise events, while quantities related to the detector response in presence of pulses are estimated from pulser events.
The pulser-based approach enables control over the timing and amplitude of the heat-injected pulses and avoids relying on the trigger settings, which would need to adapt for any change in NTD thermistor working conditions.

\begin{figure}[t!]
  \begin{center}
    \includegraphics[width=0.9\textwidth]{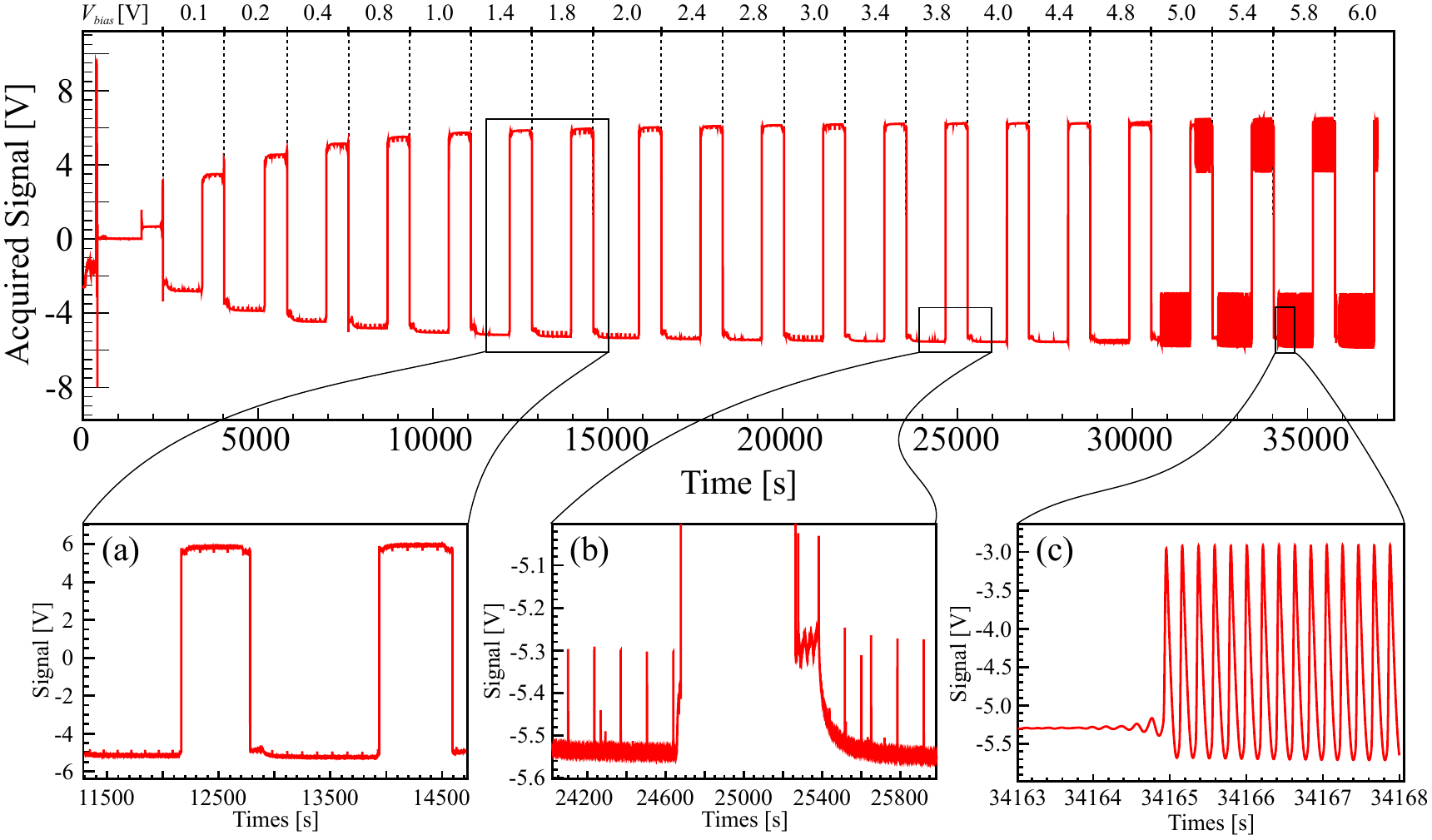}
    \caption{Continuous waveform output during a {\it{load curve measurement}} for one CUORE detector.
      For each bias configuration, EleId, the data are acquired both for negative and positive polarity.
      In zoom (a), we see two different bias configurations with data taken at negative and positive polarities; the baseline inversion is clearly visible.
      Zoom (b) shows the transient time spent in configuring the electronics, waiting for the detectors to stabilize and for performing the actual measurement for each polarity for a given EleId.
      In the last high-bias configurations, the detector output is oscillating, as reported in zoom (c).}
    \label{fig:lcwaveform}
  \end{center}
\end{figure}

From a procedural point of view, any {\it{characterization measurement}} consists of applying a sequence of electronics settings to the detectors, and to acquire, for each of these configurations, an arbitrary number of noise events and pulser events.
All the following data processing is left to the offline analysis, see sec.\,\ref{sec:lc_analysis}.
The configurable number of noise and pulser events are usually chosen larger than one, so that events that do not pass quality cuts can be discarded in the analysis, and the estimated quantities can be averaged over multiple events. The {\it{characterization measurements}} introduced before are implemented as:
\begin{itemize}
\item {\it{Load curve measurement.}} In the I-V curve, both positive and negative bias polarity configurations are measured in a wide interval of bias voltages. In this case, the main purpose is to measure $V_{NTD}$ as a function of the bias current. An example of the detector output for this type of measurement is reported in fig.\,\ref{fig:lcwaveform}.
\item {\it{Working point measurement.}} In the SNR curve, only negative polarity bias configurations are measured in a sub-range of bias voltages that includes the optimal WP.
  A large number of noise and pulser events, much more than those taken during the {\it{load curve measurement}}, are acquired in order to improve the accuracy of the measurement.
  The preamplifier output is offset such that we can maximize the signal gain of the readout chain for the negative polarity configurations.
  In this way, we can fully exploit the digitizer ADC dynamic range, increasing the precision in the reconstruction of all the pulse-related parameters.  
\item {\it{Resistance measurement.}}
The NTD thermistor resistance value for a given applied bias voltage is obtained from measuring a single point of the I-V curve. 
The NTD thermistor voltage, current (and resistance) are obtained from two configurations with opposite polarity and the same bias voltage, which is usually the one chosen for physics measurements.
\end{itemize} 
All these {\it{characterization measurements}} are usually performed in parallel on a large number of detectors.
Their duration is determined by the number of electronics configurations, the number of noise and pulser events acquired for each configuration, and by the time needed for the detectors to stabilize after an electronics configuration is applied.
For example a {\it{resistance measurement}} performed in parallel over all the CUORE detectors lasts about 3\,hours, while a {\it{load curve measurement}} performed over 3 CUORE towers takes about 12 hours.

\subsection{Measurement procedure}\label{sec:lc_processing}
The three {\it{characterization measurements}} introduced above share a similar measurement procedure in the data acquisition system, as illustrated in the following. Although this procedure was designed with CUORE detectors in mind, it can be also adapted to any cryogenic macro-calorimetric experiment using NTD thermistors as phonon sensors.
The {\it{load curve measurement}} is the most general procedure since it is partially comprised of the {\it{working point}} and {\it{resistance measurement}} procedures. 
The first step consists of adjusting the preamplifier gain and offset in order to exploit, at best, the entire ADC dynamic range to avoid saturation and to ensure that, during the measurement, the output signal takes values in a range almost symmetric around zero. 
For each point of the load curve, the measurement consists of applying the required bias voltage at negative bias polarity, waiting some time to reach a steady output, and acquiring the baseline voltage for the chosen number of noise/pulser events.
The bias polarity is then inverted and the measurement is repeated. 
The purpose of measuring the output signal at both positive and negative bias voltage polarity is to cancel the contribution of the residual input offset when measuring the average value of the output signal.
A reference heater pulse, with fixed energy of $\sim$1\,MeV, is periodically fired for the entire duration of the measurement. 
The heater pulse amplitude acts as a reference to monitor the signal-to-noise ratio, the pulse amplitude, and shape variation for every applied bias voltage.
In order to simplify the data reconstruction, {in this analysis we use only the} pulser events from the negative bias polarity, where each signal is characterized by a fast upward pulse; therefore the duration of each negative polarity step is driven by the number of pulses that are required to obtain a good estimate of their amplitude and shape parameters. Since the only use of the positive polarity step is to cancel the residual input offset, its duration is determined only by the number of {{noise events}} required. 
Once one point of the load curve is completed, the polarity is switched again to negative and a new higher value of the bias is set (see fig.\,\ref{fig:lcwaveform} - zoom (a)).
This procedure is repeated with increasing bias voltage, as shown in fig.\,\ref{fig:lcwaveform}, until a maximum bias value is reached. The maximum bias is set either to a pre-determined value or it is identified by the appearance of the oscillation effects mentioned in sec.\,\ref{sec:lc_ntd} and shown in the bottom right insertion of fig.\,\ref{fig:lcwaveform} (c).
It is worth noting that between each configuration there is a transient region of thermal origin, as shown in fig.\,\ref{fig:lcwaveform} (b).
Changing any electronics parameter, and in particular the bias voltage value, has a thermal effect on the output signal, and it can take up to a few hundreds of seconds before the signal returns to a steady condition.
These transient time windows are not associated to any EleId and the corresponding events are discarded in the reconstruction and analysis phase. 
The values of the bias voltage for each NTD thermistor are specified in a configuration file. 
Similarly, the number of steps, the duration of each step, the number of pulses in the negative polarity and the number of {noise events} to be acquired in the positive polarity, are all specified in a configuration file, making the procedure flexible and easy to adapt to the different {\it characterization measurements}. 
For each EleId, the typical number of {noise events} is between 5 and 15, while the typical number of {pulsers events} is between 10 and 30, depending on the required accuracy.

\section{Optimal working points in CUORE}\label{sec:lc_analysis}

In this section we discuss the procedures to evaluate the relevant parameters for the selection of the optimal WP, namely the NTD thermistor voltage ($V_{NTD}$), pulse amplitude (A), noise (N) and signal-to-noise ratio (SNR) as a function of the applied bias voltage. The analysis techniques presented in this section, and in particular the pulse-shape quality checks, have a fundamental role in the identification of the optimal operating conditions of the detectors. They also allow the measurement of the detector  resistances, which in turn guarantees an accurate monitoring of the stability of the detectors during data-taking.
All data shown in the first part of this section were acquired by CUORE at a base temperature  of 11.8~mK. The section concludes with a discussion of the characterization procedures applied to a wide range of temperatures, further demonstrating their versatility.

\subsection{From raw-data to characterization plots}\label{sec:rawdatatoplot}

Once a \emph{characterization measurement} has been acquired, the reconstruction of the raw data is performed through a sequence of event-based analysis modules \cite{Alduino:2016zrl}. 
The goal of the reconstruction software is to obtain all the parameters ($V_{NTD}$, SNR, A and N) as a function of the NTD thermistor current I. 
The analysis steps necessary to compute the above mentioned parameters are repeated for each EleId, same bias current and opposite polarities.
In this section, we plot $V_{NTD}$ as a function of I instead of the canonical way to represent the load curve (see fig.\,\ref{fig:boloBias}); this allows us to easily superimpose it to the curves for the A, N and SNR variables, having a common axis I. 
In fig.\,\ref{fig:lcPlot}, there is an example for each of the curves built for a single NTD thermistor. 
In the following, we will explain how all the mentioned quantities, associated with the NTD thermistor static and dynamic response, are measured and calculated for each point of the load curve. In general, event-based variables, such as baseline and noise, are averaged to calculate the specific quantities of our interest for each bias and polarity configuration. 

As mentioned before, the NTD thermistor voltage is small and it requires amplification. We refer with $V_{bsl}$ as the voltage signal read after the amplification stage. That is defined as:
\begin{equation}\label{eq:Vbsl}
V_{bsl}^{\pm} = A_V \cdot (V_{off} \pm   V_{NTD})
\end{equation}
where $V_{off}$ is the residual input-referred offset of the electronics, $A_V$ is the gain of the amplification stage, and the $\pm$ sign corresponds to the bias polarity. 
$V_{bsl}^{\pm}$ are evaluated as the average of all the baseline values of events with the same EleId for each bias polarity. The baseline is an event-based variable and it is estimated from a linear fit of the samples in the event window. For each EleId and polarity, the reconstruction software selects a clean sample of noise events that do not exhibit drifting baseline, particle induced pile-up, etc; for these events, the average of the baseline values is calculated, resulting in $V_{bsl}^{+}$ and $V_{bsl}^{-}$.\\
The NTD thermistor voltage $V_{NTD}$ and current I are then calculated for each EleId:
\begin{equation}\label{eq:vbol_LC}
V_{NTD}= \frac{V_{bsl}^+ - V_{bsl}^-}{2 \cdot A_V} \quad , \quad
I = \frac{V_{bias} - V_{NTD}}{2\cdot R_L}
\end{equation}
 Each NTD thermistor resistance is obtained from the ratio of $V_{NTD}$ and I for each EleId.

The noise level N of each EleId is evaluated exploiting the optimum filter \cite{Gatti:1986cw}.
N is computed as the integral of the average noise power spectrum of the {noise events}, weighting each frequency based on its contribution to the average pulse, resulting in the resolution of the optimum filter.
An alternative and simpler method consists of computing for each {noise event} the standard deviation of the NTD thermistor output samples; the average of these values is taken as N for each EleId. 
 We preferred to use the first method, although more complicated, since it  suppresses frequencies irrelevant to the detector signal bandwidth. 

In order to compute the pulse amplitude A for each EleId, we utilize {pulser events}. In this case, the event-based variables are the amplitude and pulse shape parameters of the heater pulse.  
The amplitude estimator A for each EleId is evaluated as the amplitude of the average pulse, which is built by averaging all the {pulser event} waveforms. The pulse amplitude is defined as the difference between the maximum of the pulse and the baseline value in the region before the trigger. 
The use of the average pulse helps to limit the effect of incoherent noise in the identification of the pulse maximum. 
The detailed description of the pulse shape parameters and the distortion indicator is provided in sec.\,\ref{sec:lc_wp}.

Eventually, the SNR is computed as the ratio of the amplitude and noise estimators for each EleId.

\subsection{Identification of the optimal working point}\label{sec:lc_wp}
From the analysis of the curves constructed for each detector in a {\it{working point measurement}}, the optimal WP is chosen following two basic requirements: maximize the signal-to-noise ratio and avoid pulse deformation. 
An automated algorithm was developed to find the optimal bias voltage for each detector.

The shape of the heater pulses at the different bias voltages is chosen as a figure of merit for identifying possible unstable conditions. 
The detector response for CUORE detectors is modeled with an empirical function whose Laplace transform has 4 real negative poles and one negative zero in the complex plane, and the pulse is considered distorted if a pair of poles departs significantly from the real axis, becoming complex conjugate~\cite{Nutini:2019jzm}. The pulse shape in the time domain results in a combination of exponentials; the exponents can be real or imaginary, depending if the poles have real and imaginary parts. The poles can be correlated to physical quantities of the detector's thermal and electrical circuits; in particular, the real part of the poles corresponds to the inverse of the typical decay time constants of the exponentials \cite{Nutini:2020draft}.

A fit of the heater pulses with the empirical template is performed and a global shape parameter S is constructed from the fit results.
Given a pole P in the complex plane, S is defined as:
\begin{equation}
S = \frac{|Im(P)| - |Re(P)|}{|P|} 
\end{equation}{}
and its value changes with the applied bias voltage; this is correlated with the ability to reconstruct and predict the pulse deformation. 
For low bias values, the pulse shape is well-described by one time constant on the rising edge and three time constants on the falling edge, thus the 4 real poles detector response model is satisfactory and the parameter S is negative.
Increasing the applied bias, the non-linear effects from the electro-thermal feedback start to be relevant and the pulse starts showing a damped-oscillation shape.
These configurations are characterized by a positive value of S, meaning the value of the imaginary term of the pair of complex conjugate poles of the detector response function is higher than its real term, thus the oscillatory part will dominate the shape of the falling edge of the pulse. 
When the configuration corresponding to the maximum of SNR is in a region of high bias, the pulse shape can become slightly deformed.
A threshold on the S parameter can then be set to identify bias configurations leading to deformed pulses.
The threshold is typically around S\,=\,0, however its exact value can be optimized based on the specific experimental configuration and can be slightly different for measurements at different base temperatures.
For CUORE data at 11.8 mK, the threshold is set as S\,$\leq$\,-0.2. 
If the maximum SNR occurs in correspondence of a deformed pulse, the optimal working point is chosen among the lower bias values on the load curve applying the same shape checks.
Eventually the optimal WP is chosen, among the points with no deformation, as the point with the highest SNR.

\begin{figure}[t!]
\centering
\includegraphics[width=\textwidth]{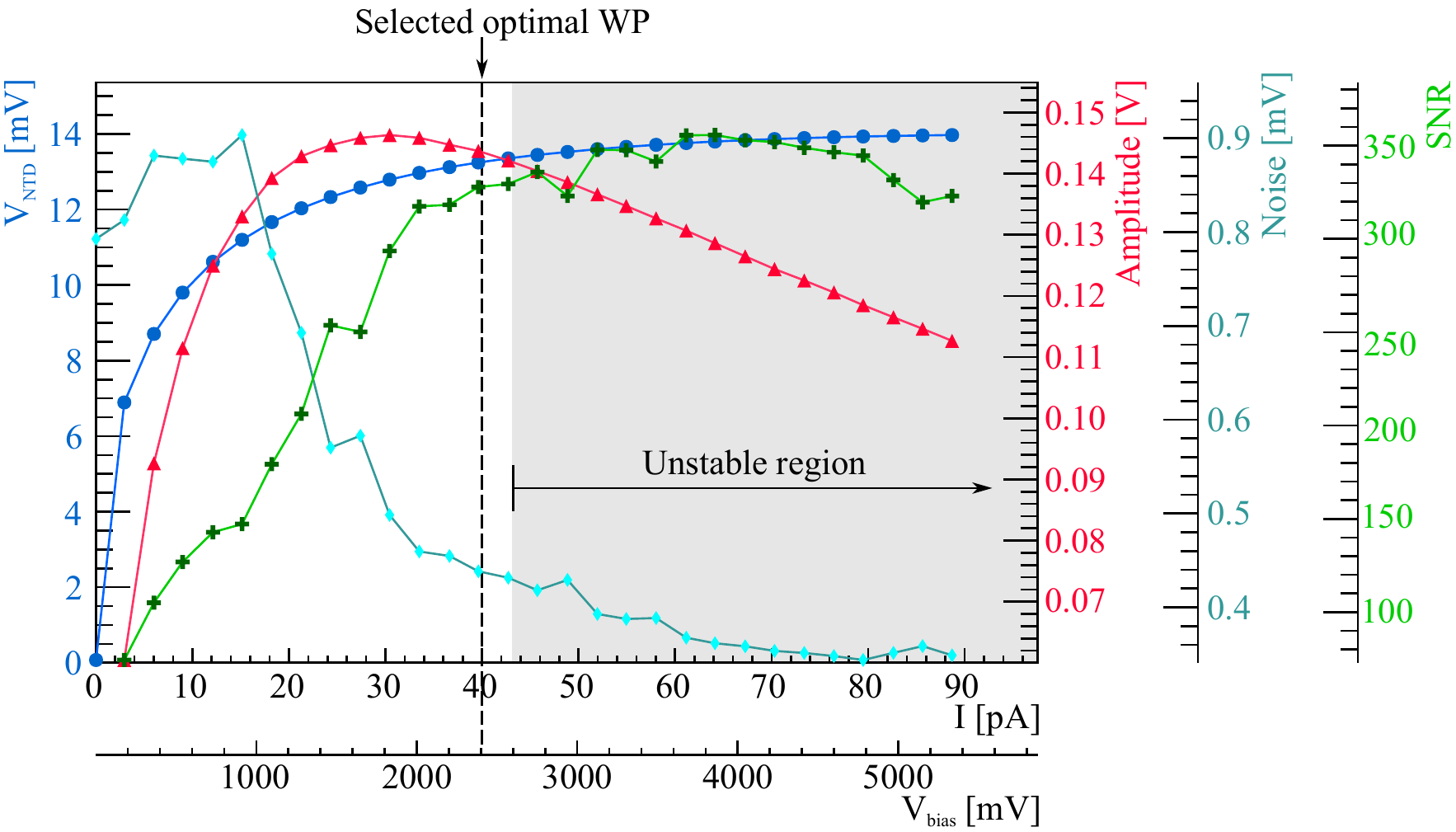}
\caption{Characterization curves produced in a {\it{working point measurement}} for one CUORE detector.
  The black dashed line indicates the point corresponding to the bias chosen as the optimal WP.} \label{fig:lcPlot}
\end{figure}

\begin{figure}[t!]
\centering
\includegraphics[width=\textwidth]{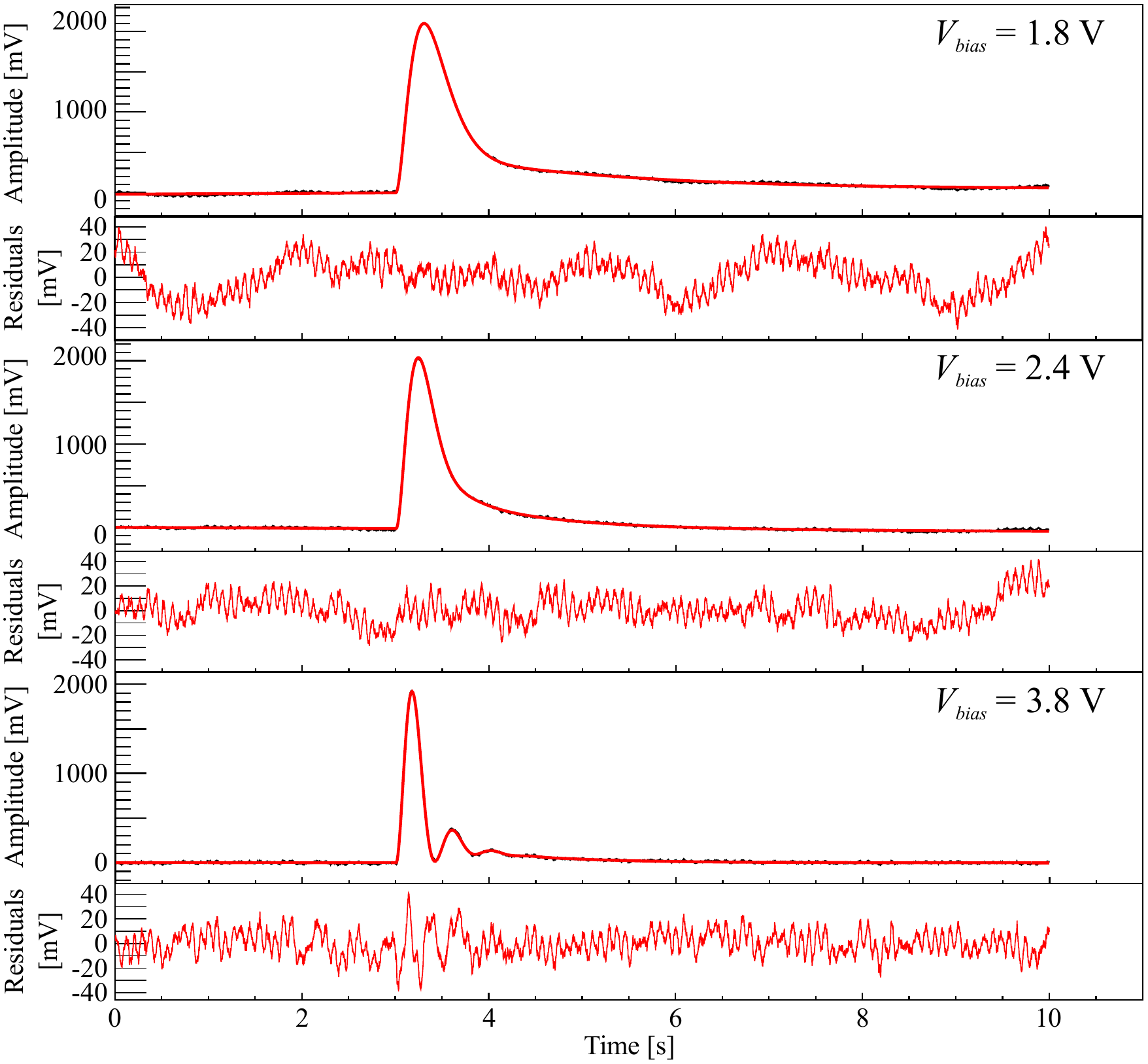}
\caption{Heater pulses, their effective fit and the fit residuals, for three different bias configurations acquired in the {\it{working point measurement}}, for the same CUORE detector whose I-V, SNR and other plots are reported in fig.\,\ref{fig:lcPlot}.
  [Top] $V_{bias}$ = 1.8 V, maximum pulse amplitude.
  [Center] $V_{bias}$ = 2.4 V, selected optimal WP.
  [Bottom] $V_{bias}$ = 3.8 V, maximum SNR. \cite{Nutini:2019jzm}}
\label{fig:exampleLC_wp}
\end{figure}

The characterization curves produced for one CUORE detector from a {\it{working point measurement}} are shown in fig.\,\ref{fig:lcPlot}.
The corresponding heater pulses at three different bias voltages (under-biased, optimal, over-biased) for the same detector are shown with the effective fit superimposed in fig.\,\ref{fig:exampleLC_wp}. 
In fig.\,\ref{fig:lcPlot}, the pulse amplitude reaches its maximum for $V_{bias}$ = 1.8 V and has an ideal shape; however the SNR is not maximized in this configuration, due to higher noise N for lower bias voltage.
On the contrary, the SNR is maximized for $V_{bias}$ = 3.8 V; the position of this point on the I-V curve is far from the linear region and the pulse shows a strong damped-oscillation shape.
Therefore the working point $V_{bias}$ = 2.4 V is chosen as a compromise between having a high SNR and non deformed pulse. 
A summary of the parameters ($V_{bias}$, A, SNR, S) for the three configurations corresponding to the pulses in fig.\,\ref{fig:exampleLC_wp} is reported in tab.\,\ref{tab:pulses_wp}.

In general, with the addition of the pulse shape condition, the SNR at the chosen WP is only slightly lower than the maximum SNR (in the example shown in tab.\,\ref{tab:pulses_wp}, the SNR reduction is $\sim$\,8\%).
The choice of the WP in a region of stable response of the NTD thermistor is as important as ensuring a high SNR.
The former condition is fundamental to maintaining a correct operation and a uniform response of the NTD thermistor in the energy range of interest. 
A stable dynamic behavior of the NTD thermistor helps in the overall data processing, in particular when we apply the Optimal Filtering technique~\cite{Gatti:1986cw} for the pulse amplitude reconstruction. This procedure assumes a uniform pulse template for a wide energy range; if there is a residual pulse shape dependence on energy, the algorithm reduces its performance in terms of energy resolution.

\begin{table}[t!]
    \centering
    \begin{tabular}{r|l|l|l}
        Bias Voltage ($V_{bias}$)      & 1.8 V  & 2.4 V & 3.8 V\\
        Pulse Amplitude (A)          & 0.148 V & 0.144 V  & 0.130 V  \\
        Signal-to-noise Ratio (SNR)         & 300 mV/mV & 330 mV/mV & 360 mV/mV \\
        Shape parameter (S)              & -0.35 & -0.20 & 0.05  \\
    \end{tabular}   
    \caption{Summary of the parameters for the three load curve configurations identified by the three pulses in fig.\,\ref{fig:exampleLC_wp}. The parameters in the second column correspond to the maximum pulse amplitude point; the third column, the selected optimal WP; and the fourth column, the maximum SNR point.}
    \label{tab:pulses_wp}
\end{table}

\subsection{Detector response at the optimal working point} \label{sec:wp_check}
\begin{figure}[t!]
\centering
\includegraphics[width=0.9\textwidth]{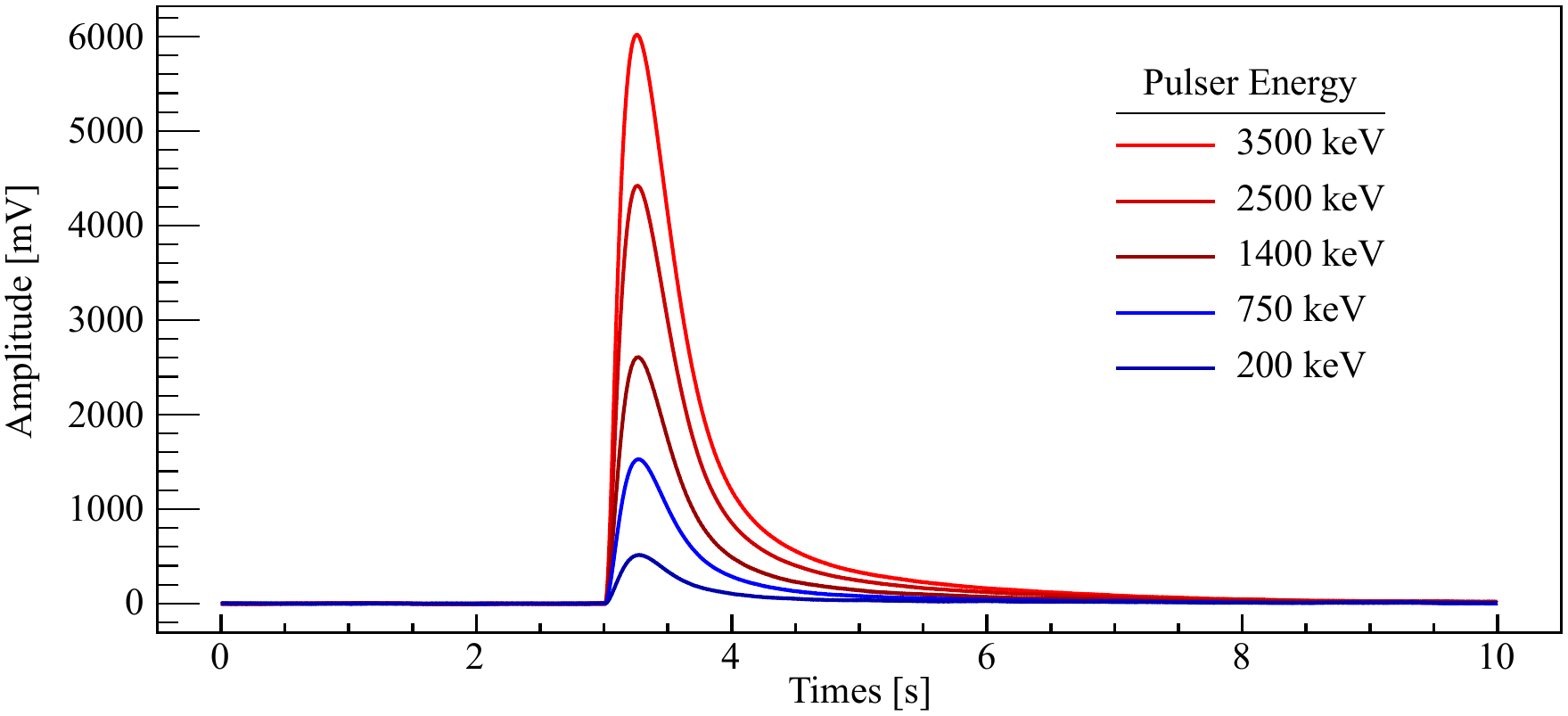}
\caption{Heater {average} pulses at different energies for the same CUORE detector as in fig.\ref{fig:exampleLC_wp}. The NTD thermistor is operated at the optimal WP, which ensures a uniform pulse shape with energy.}
\label{fig:npulsesCompare}
\end{figure}

\begin{figure}[t!]
\includegraphics[width=0.95\textwidth]{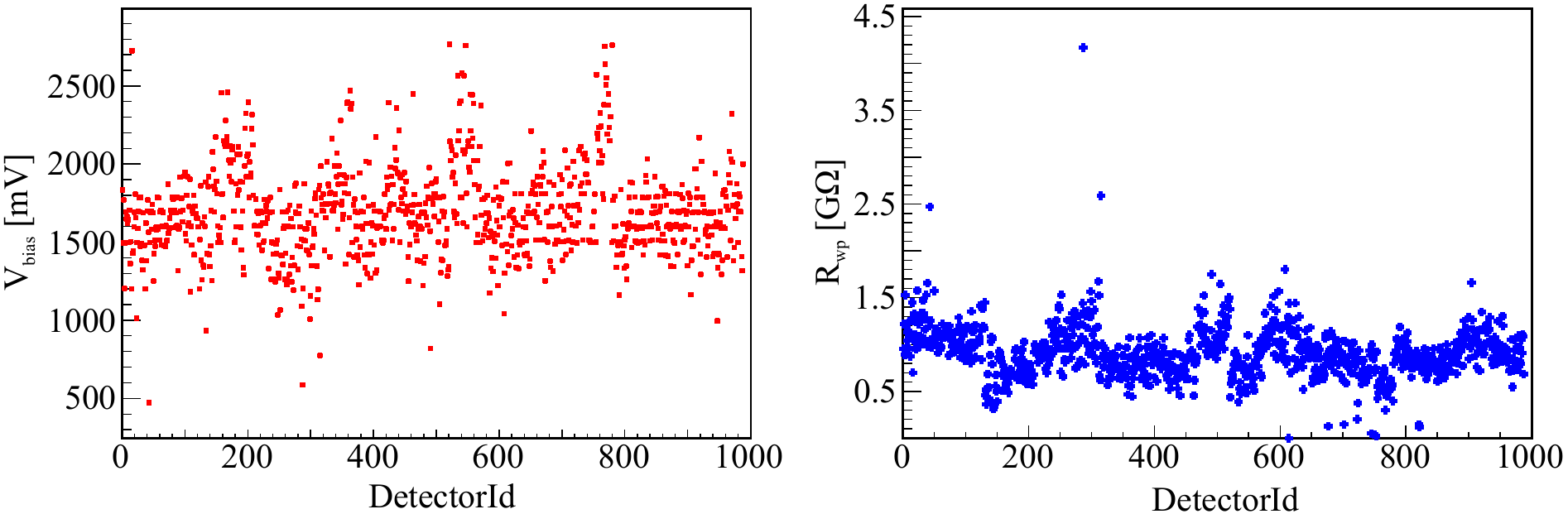}
\caption{[Left] Distribution of the bias voltages applied to the NTD thermistor as a function of the CUORE {{DetectorId}}. [Right] Distribution of the NTD thermistor resistances at the optimal WP as a function of the CUORE {{DetectorId}}. Both plots in the figure correspond to 11.8 mK base temperature.}
\label{fig:wp11mK}
\end{figure}

After having set the optimal working points, a further measurement is performed in which pulser events of various energies (200 keV - 3.5 MeV) are injected in the detectors. The purpose of this measurement is to verify that the pulses preserve a proper shape even at energies lower/higher than that of the pulser used during the {\it{working point measurement}}. The working point is accepted if the parameter S described in sec.\,\ref{sec:lc_wp} is consistent over the scanned energy range for a given detector. For the detectors with varying S parameters, the WP is adjusted manually. In CUORE this condition occurred in $\sim$3\% of the detectors, which in most cases corresponded to noisy detectors for which the SNR curve was reconstructed with poor accuracy in the {\it{working point measurement}}. 
The final optimal WP exhibits a uniform pulse shape over the pulser energy range (200 keV - 3.5 MeV) , see fig.\,\ref{fig:npulsesCompare}.\\
The optimal WP bias voltages applied to the CUORE detectors at an operating temperature of 11.8 mK are reported in fig.\,\ref{fig:wp11mK}[left]. The voltages cluster mainly between 1.5 - 2.0 V.
This corresponds to a range of $R_{wp}$ $\sim$ 0.6 - 1.2 $G\Omega$, see fig.\,\ref{fig:wp11mK}[right].

\begin{figure}[ht!]
\centering
\includegraphics[width=\textwidth]{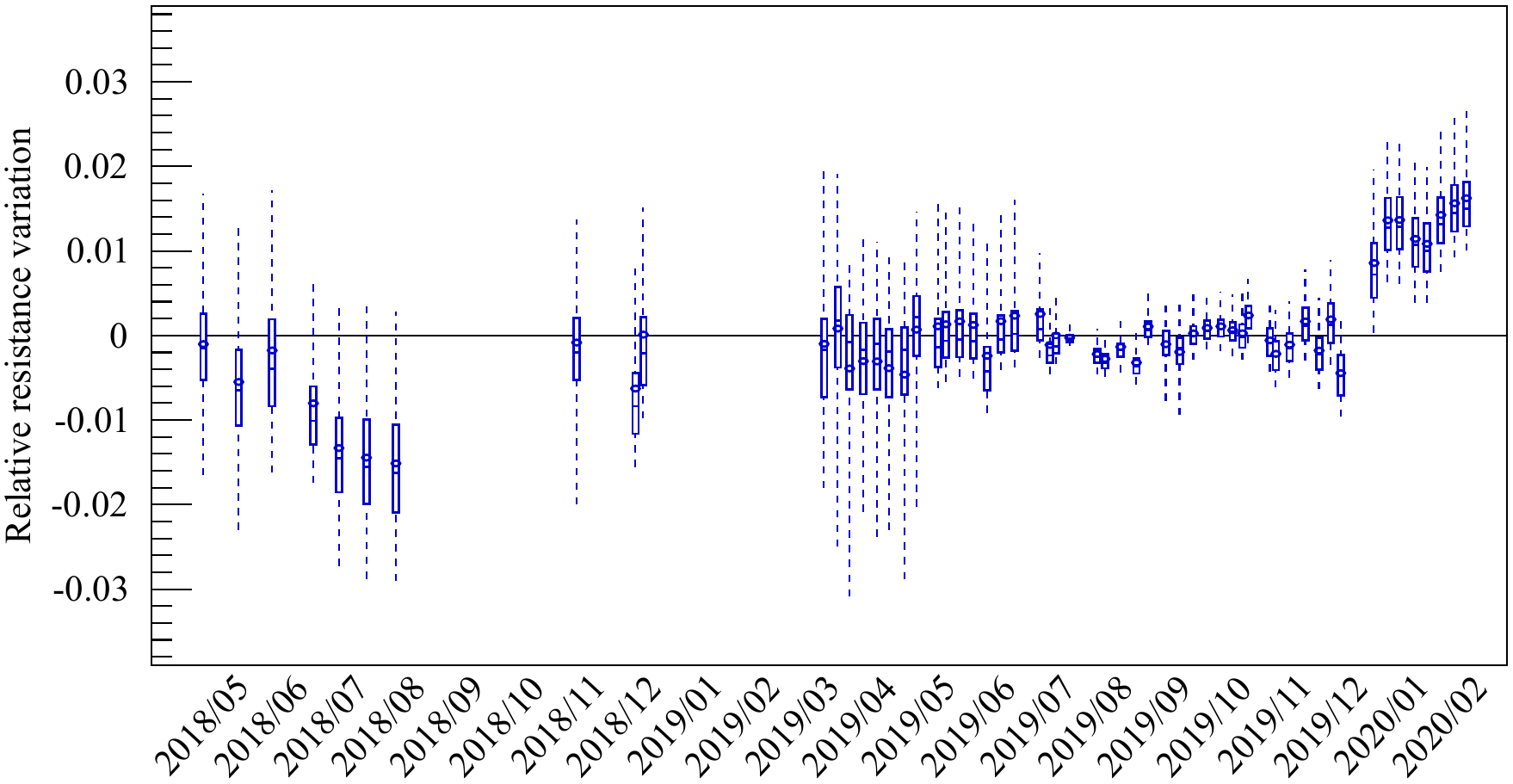}
\caption{CUORE resistance stability over 2 years of operation at 11.8 mK. 
Each element in the plot corresponds to the distribution of the relative variation of $R_{wp}$ taken during a single {\it{resistance measurement}} for all the CUORE detectors. {A given {\it{resistance measurement}} is taken as reference ($R_{wp}^{ref}$) for each temperature; then the difference between the reference and each single measurement ($\Delta R$\,=\,$R_{wp}-R_{wp}^{ref}$) is calculated and normalized by the reference value, in order to evaluate the relative variation $\Delta R / R_{wp}^{ref}$ for each channel.}
  The circle corresponds to the mean of the distribution and the line close to that is the median. The solid box around the mean indicates the 50\% of the distribution of the data.
  The dashed lines extending vertically from the boxes indicate the width of 95\% of the data distribution.}
\label{fig:exampleR_wp_stability}
\end{figure}

During the standard CUORE data-taking, the effective NTD thermistor resistance at the optimal WP ($R_{wp}$) is measured for each detector a few times per month in order to monitor the devices' thermal stability over time; this is the {\it{resistance measurement}} described in sec.\,\ref{sec:daq}. In fig.\,\ref{fig:exampleR_wp_stability} the stability of the NTD thermistor working point resistance for all the detectors during the CUORE data taking at 11.8 mK is reported. We evaluated the relative variation with time of the $R_{wp}$ for each detector compared to a corresponding reference measurement (from August 2019). Time intervals with no points correspond to periods of cryogenic maintenance and no data-taking. The data are more sparse in 2018, since {\it{resistance measurements}} were performed bi-weekly or monthly in that period. Since 2019, we performed these measurements weekly to allow for a continuous and more accurate monitoring of the thermal stability of the detectors and the overall system. The average resistance variation shows a small jump in the data between late 2019 and early 2020, that is correlated with a slight change in the operating temperature. For the same applied bias voltages on the NTD thermistors, a lower temperature led to larger resistance values. 

\subsection{Measurements at different temperatures}\label{sec:temperature_scan}
\begin{figure}[ht!]
\centering
\includegraphics[width=\textwidth]{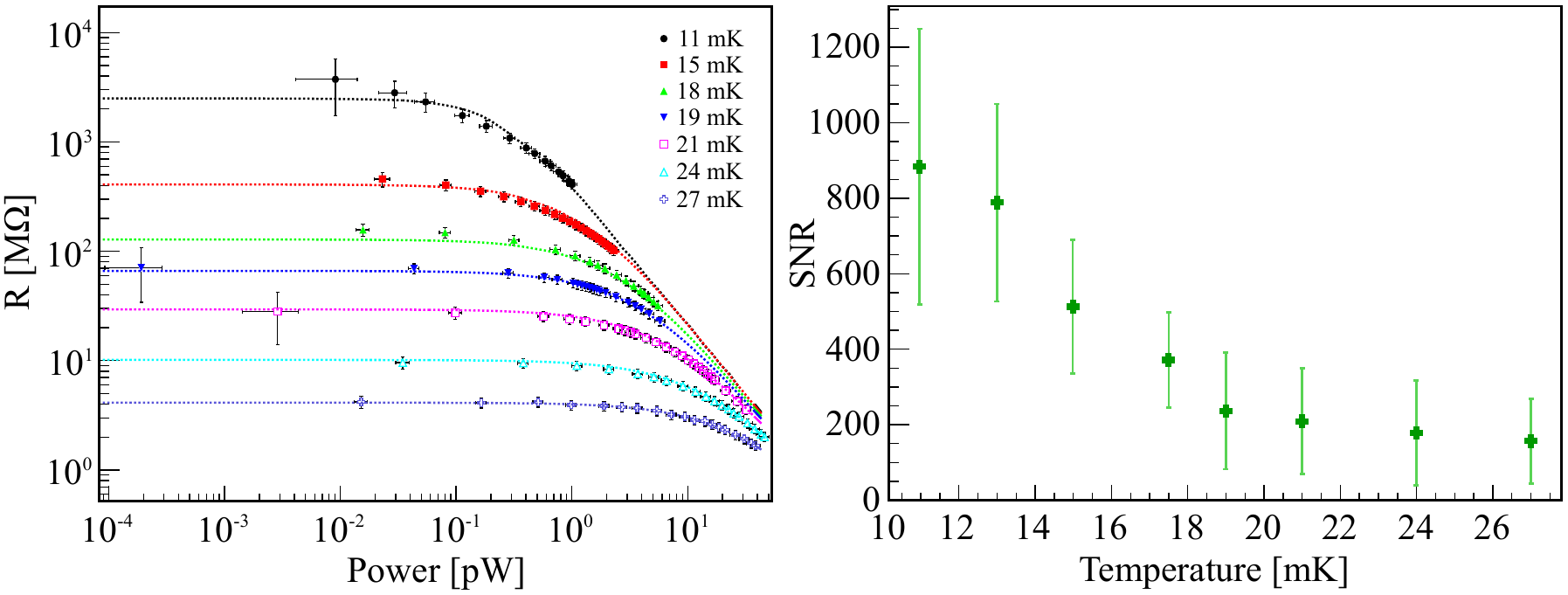}
\caption{[Left] R-P curves of one CUORE NTD thermistor for several values of base temperature, from 11 mK to 27 mK.  {The superimposed dashed lines are graphical interpolations of the data points for each temperature.}
  [Right] Variation of the average SNR evaluated over a subset of detectors for several values of base temperature.}
\label{fig:boloLC_T}
\end{figure}

A dedicated set of {\it{load curve measurements}} at temperatures ranging from 11 mK to 27 mK base temperature was performed for CUORE.
The data were acquired and analyzed with the procedures described above. 
An instructive visualization of these measurements is provided by the R-P curve (resistance vs. injected power), as reported in fig.\,\ref{fig:boloLC_T}[left] for one CUORE detector.
These curves show the combined effect of variation of the NTD thermistor base resistance with temperature, R($T$;\,$P$=0), and its dependence on the power dissipation due to the electro-thermal feedback for increasing bias voltages $V_{bias}$.
Curves with lower base resistance correspond to higher base temperatures.
Incidentally, we {deduce that the lower injected power used to bias the thermistor corresponds to a minimum bias current of a few pA; this proves that} the amplifier differential input current of a few tens of fA (see sec.\,\ref{sec:ele}) has a completely negligible effect.

To inspect the temperature dependence of the dynamic behavior of the detectors, we evaluated, for each temperature, the distribution of the SNR at the optimal WP for the CUORE detectors. Each point in fig.\,\ref{fig:boloLC_T}[right] corresponds to the average of this distribution at a given temperature, while the
vertical bar represents the standard deviation of the mean. {In particular, in the figure the SNR is evaluated using the higher pulser amplitude (at $\sim$ 3.5 MeV) and the filtered noise level N.}
We observed that in the measured energy and temperature ranges,  the SNR improves for lower temperatures.
We ascribed this behavior to the larger absorber internal gain for lower temperatures, which over-compensates the increase in the noise level due to larger NTD thermistor resistance values. The absorber internal gain, in the case of cryogenic calorimeters, is considered the conversion factor from deposited energy to the amplitude of the thermal pulse. The absorber gain is inversely proportional to the heat capacity C, with C $\propto$ $T^3$ in dielectric materials such as TeO$_2$. Therefore, a lower heat capacity C, obtained for lower base temperatures, leads to a larger signal amplitude for a fixed energy release.

From this study, it was possible to identify the best operating temperature of the CUORE
setup.
A more detailed discussion on the CUORE temperature studies and detector thermal response optimization will be part of a future work.

\section{Conclusions}
A dedicated procedure for performing automatic load curve measurements on a large number of cryogenic calorimeters read by NTD thermistors has been developed.
The specific algorithm analyzes the load curve data and identifies the optimal bias voltage (optimal WP) for each NTD thermistor.
Furthermore, the addition of the information related to the pulse shape dependence with the applied bias constitutes an upgrade of the standard approach of choosing optimal WP bias voltages corresponding to either only the maximum amplitude or maximum SNR.

This is the first time that an automatic procedure is utilized for setting the optimal WP for almost one thousand NTD thermistors coupled to TeO$_2$ crystals and can be reproduced for load curve data acquired at any base temperature. 
This was crucial to find the best operating settings of the CUORE detector. 
Other cryogenic experiments employing macro-calorimeters coupled with NTD thermistors could profit from the highly automated system presented in this paper, for easing the procedures to optimize the detectors' operating settings.
{The optimal WP search algorithms presented in this paper were developed having in mind the specific needs of CUORE, however other algorithms could be easily accommodated in the analysis framework in order to meet the needs of other experiments.
For example the criteria for the choice of the optimal WP could be based also on parameters that were not relevant or applicable in the case of CUORE, such as the pile-up discrimination~\cite{CUPIDInterestGroup:2020rna} or the pulse-shape-based particle discrimination capability~\cite{Bandac:2021qco, Beeman:2012ci} of the detectors.} 

\section*{Acknowledgment}

This work was sponsored by the Istituto Nazionale di Fisica Nucleare (INFN).
In addition we would like to acknowledge the support from the DOE Office of Science, Office of Nuclear Physics.
The authors thank the CUORE Collaboration, the directors and staff of the Laboratori Nazionali del Gran Sasso.

\bibliographystyle{elsarticle-num}
\bibliography{main}

\end{document}